\documentclass{article} % For LaTeX2e
\usepackage{iclr2017_workshop,times}
\usepackage{hyperref}
\usepackage{url}
\usepackage{graphicx}
\usepackage{booktabs}       % professional-quality tables
\usepackage{amsfonts}       % blackboard math symbols
\usepackage{nicefrac}       % compact symbols for 1/2, etc.
\usepackage{microtype}      % microtypography

\title{Fast Wavenet Generation Algorithm}

\author{Tom Le Paine$^1$, Pooya Khorrami$^1$, Shiyu Chang$^2$, Yang Zhang$^1$,\\
{\bf Prajit Ramachandran$^1$, Mark A. Hasegawa-Johnson$^1$ \& Thomas S. Huang$^1$}\\
\\
$^1$University of Illinois at Urbana-Champaign, IL 61801, USA \\
\texttt{\small \{paine1, pkhorra2, yzhan143, prmchnd2, jhasegaw, t-huang1\}@illinois.edu} \\
\\
$^2$IBM Thomas J. Watson Research Center, NY 10598, USA \\
\texttt{\small shiyu.chang@ibm.com}
}

\begin{document}

\maketitle

\begin{abstract}
This paper presents an efficient implementation of the Wavenet generation process called Fast Wavenet. Compared to a na\"{i}ve implementation that has complexity O$(2^L)$ ($L$ denotes the number of layers in the network), our proposed approach removes redundant convolution operations by caching previous calculations, thereby reducing the complexity to O$(L)$ time.  Timing experiments show significant advantages of our fast implementation over a na\"{i}ve one. While this method is presented for Wavenet, the same scheme can be applied anytime one wants to perform auto-regressive generation or online prediction using a model with dilated convolution layers. The code for our method is publicly available.
\end{abstract}

%%%%%%%%%%%%%%%%%%%%%%%%%%%%%%%%%%%%%%%%%%%%%%%%%%%%
\section{Introduction}
Wavenet \citep{oord2016wavenet}, a deep generative model of raw audio waveforms, has drawn a tremendous amount of attention since it was first released. It changed existing paradigms in audio generation by directly modeling the raw waveform of audio signals. This has led to state-of-the-art performance in text-to-speech and other general audio generation settings including music.  

Wavenet models the conditional probability via a stack of dilated causal convolutional layers for next-sample audio generation given all of the previous samples. At training time, since the audio samples for all timestamps are known, the conditional predictions can be naturally made in parallel. However, when generating audio using a trained model, the predictions are sequential. Every time an output value is predicted, the prediction is then fed back to the input of the network to predict the next sample.

\begin{figure}[ht]
  \centering
  \includegraphics[width=0.9\textwidth]{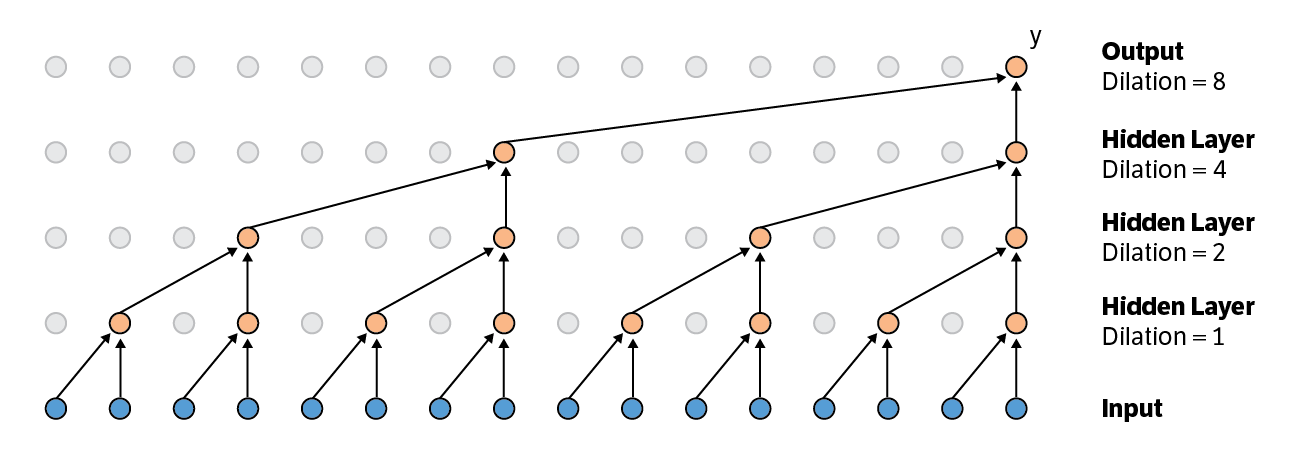}
  \caption{Na\"{i}ve implementation of generation process. Notice that generating a single sample requires $O(2^L)$ operations.}
  \label{fig:naive_gen}
\end{figure}

In Figure \ref{fig:naive_gen}, we show a toy Wavenet network used to compute the value of a single output node (A dynamic visualization can be found at DeepMind’s blog post\footnote{\url{https://deepmind.com/blog/wavenet-generative-model-raw-audio}}). The input nodes (blue) are the leaves of the tree, and the output is the root. The intermediate computations are the orange nodes. The edges of the graph correspond to matrix multiplications.  Since the computation forms a binary tree, the overall computation time for a single output is O$(2^L)$, where $L$ is the number of layers in the network.  When L is large, this is extremely undesirable. 

This work fills a missing piece of the original Wavenet paper by providing an efficient implementation for audio generation. The main ingredient of the proposed approach is that we store necessary intermediate computations. 

The na\"{i}ve implementation in Figure \ref{fig:naive_gen} recomputes many variables that have already been computed for previous samples.  Note that, though we call the implementation in Figure \ref{fig:naive_gen} ``na\"{i}ve'', it is the implementation used in previous open source distributions \footnote{\url{https://github.com/ibab/tensorflow-wavenet}}.  By caching previous computations, we can reduce the computational complexity of generating a single output to O$(L)$. We call our efficient implementation: Fast Wavenet \footnote{\url{https://github.com/tomlepaine/fast-wavenet}}.

While we present this fast generation scheme for Wavenet, the same scheme can be applied anytime one wants to perform auto-regressive generation or online prediction using a model with dilated convolution layers. For example, the decoder in ByteNet \citep{kalchbrenner2016neural} performs auto-regressive generation using dilated convolution layers, therefore our fast generation scheme can be applied.

%%%%%%%%%%%%%%%%%%%%%%%%%%%%%%%%%%%%%%%%%%%%%%%%%%%%
\section{Fast Wavenet}
The key insight to Fast Wavenet is the following: given a specific set of nodes in the graph, we will have sufficient information to compute the current output. We call these nodes the recurrent states in reference to recurrent neural networks (RNNs) \citep{graves2013generating}. An efficient algorithm can be implemented by caching these recurrent states, instead of recomputing them from scratch every time a new sample is generated.

\subsection{A Graphical Illustration}
\begin{figure}[t]
  \centering
  \includegraphics[width=0.9\textwidth]{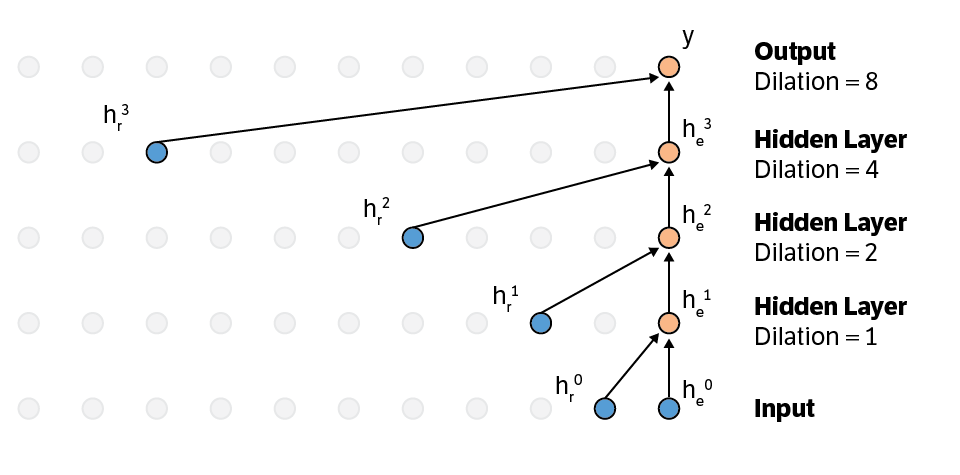}
  \caption{Simplified computation graph produced by our Fast Wavenet method. Now for a single output, the computational complexity is $O(L)$ where $L$ is number of layers in the network.}
  \label{fig:simple_graph}
\end{figure}

The graph displayed in Figure \ref{fig:simple_graph} illustrates the idea of the recurrent states. This graph, like the one in Figure \ref{fig:naive_gen}, shows how a single output sample is generated except now it is in terms of the pre-computed ("recurrent") states. In fact, upon closer inspection, the reader will notice that the graph shown in Figure \ref{fig:simple_graph} looks exactly like a single step of a multi-layer RNN. For some given time $t$, the incoming input sample $(h^{0}_e)$ can be thought of as the "embedding" input and is given the subscript 'e'. Similarly, the recurrent states are given subscript 'r'. Since these recurrent nodes have already been computed, all we need to do is cache them using a queue. From Figure \ref{fig:simple_graph}, we see by using cached values, the generation process now has complexity $O(L)$.

\begin{figure}[t]
  \centering
  \includegraphics[width=0.9\textwidth]{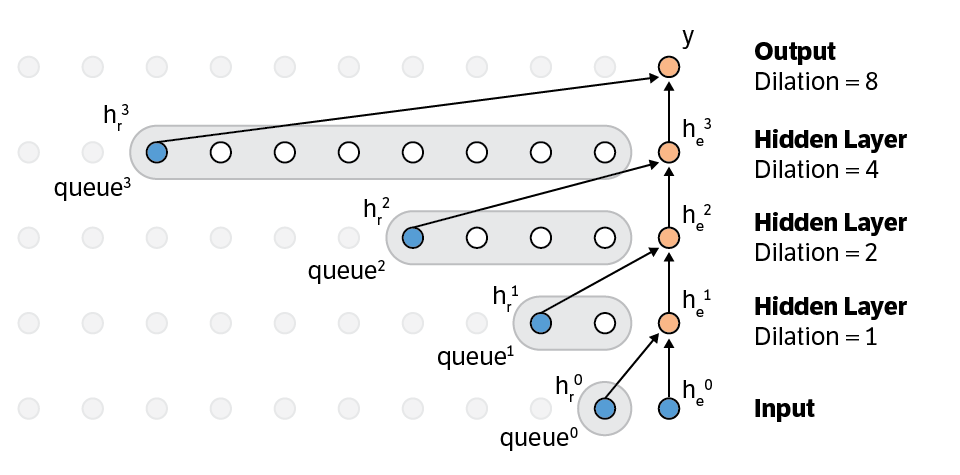}
  \caption{ The caching scheme for efficient generation. Due to dilated convolutions, the size of the queue at the $l^{th}$ hidden layer is $2^l$. }   \label{fig:what_to_store}
\end{figure}

However, it should be noted that, due to the dilated convolutions, outputs at each layer will depend on the stored recurrent states computed several time steps back, not the immediate predecessors. Thus, we can use a first-in-first-out queue in each layer to cache the recurrent states that are yet to be used. The number of states cached at each layer is determined by the dilation value of the layer. We provide an example in Figure \ref{fig:what_to_store}. For the first hidden layer, it has a dilation value of 1, therefore the queue below this layer, denoted $(\texttt{queue}^{0})$ in the figure, only needs to keep track of 1 value. On the other hand, the output layer has a dilation value of 8, which means the queue housing the previous recurrent states below this layer, denoted as $(\texttt{queue}^{3})$, is size 8. 

\subsection{Algorithm}
Our algorithm has two main components:
\begin{itemize}
\item Generation Model
\item Convolution Queues
\end{itemize}

\begin{figure}[t]
  \centering
  \includegraphics[width=0.9\textwidth]{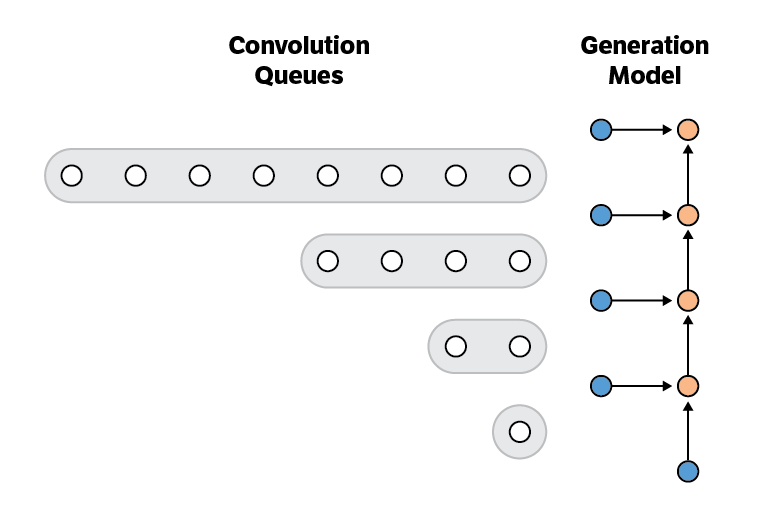}
  \caption{Two main components of our algorithm: generation model and convolution queues.  The generation model can be viewed as a single step of a multi-layer RNN, where the recurrent inputs are fed by timely updated convolution queues.}
  \label{fig:components}
\end{figure}

They are shown visually in Figure \ref{fig:components}. As we described previously, the generation model resembles and behaves like a single step of a multi-layer RNN. Specifically, it takes in the current input along with a list of recurrent states and produces the current output, along with the new recurrent states. The convolution queues store the recurrent states and are updated when the new recurrent states are computed.

To generate audio, we first initialize the generation model using the weights from a pre-trained Wavenet model. Next, we initialize the convolution queues by setting all of their recurrent states to zeros. Then, when generating each output sample, we perform the following steps:
\begin{itemize}
\item Pop Phase
\item Push Phase
\end{itemize}

\begin{figure}[t]
  \centering
  \includegraphics[width=0.85\textwidth]{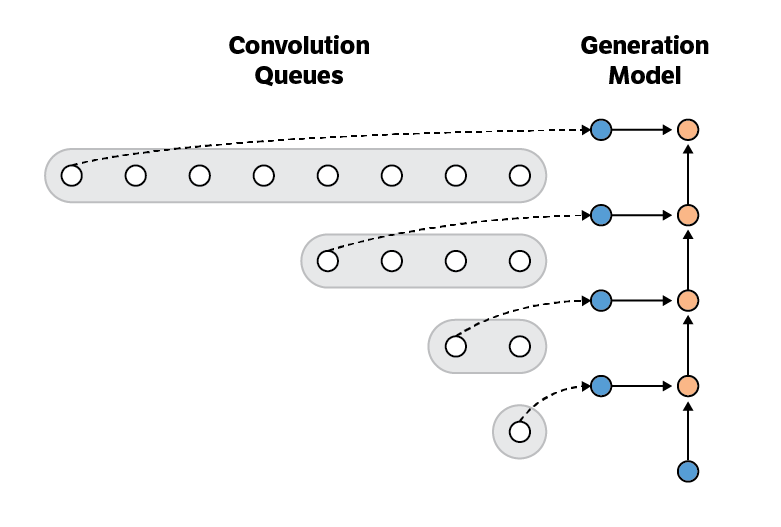}
  \caption{Pop phase: the recurrent states are popped off of each convolution queue and fed as input (blue dots) into the corresponding location of the generation model. These values along with the current input (bottom blue dot) are used to compute the current output and the new recurrent states (orange dots).}
  \label{fig:pop}
\end{figure}

\begin{figure}[t]
  \centering
  \includegraphics[width=0.85\textwidth]{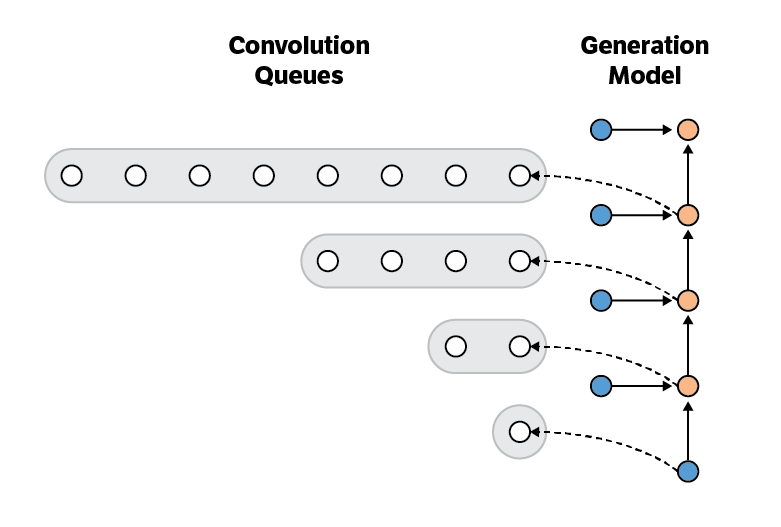}
  \caption{Push phase: the new recurrent states (orange dots) are pushed to the back of their respective convolution queues.}
  \label{fig:push}
\end{figure}

During the pop phase, the first recurrent state is popped off of each convolution queue and fed into the corresponding location of the generation model. These values along with the current input are used to compute the current output and the new recurrent states. This process is illustrated in Figure \ref{fig:pop}. Once the new recurrent states have been computed, they are then pushed into their respective queues during the push phase, as shown in Figure \ref{fig:push}.

%%%%%%%%%%%%%%%%%%%%%%%%%%%%%%%%%%%%%%%%%%%%%%%%%%%%
\section{Complexity Analysis}
In this section, we demonstrate the advantage of our Fast Wavenet algorithm over a na\"{i}ve implementation of the generation process both theoretically and experimentally.

\subsection{Theoretical Analysis}
Here we briefly summarize the complexity of both the na\"{i}ve and proposed simplified implementations. In terms of computational complexity, the simplified implementation requires $O(L)$, whereas a previous implementation of the algorithm in Figure \ref{fig:naive_gen} requires $O(2^L)$.

In terms of space complexity, the simplified implementation needs to maintain $L$ queues, which altogether occupy $O(2^L)$ additional space. On the other hand, the na\"{i}ve implementation needs to store intermediate hidden outputs. Assuming the intermediate results of the lower hidden layer will be erased after those of the higher layer are computed, the additional space required by the na\"{i}ve implementation is also $O(2^L)$. In short, the proposed implementation saves computational complexity dramatically without compromising space complexity.

It is also worth mentioning that the proposed implementation scales well to more general architectures. For an architecture with filter width $w$, and convolution rate of the $l$th layer $r^l$, assuming $r\geq w$,the proposed implementation requires $O(wL)$ computation and $O((w-1)r^L)$ additional space to generate a new sample, while the na\"{i}ve version requires $O(w^L)$ and $O(w^{L-1})$ respectively. The computational complexity differs greatly, but the space complexity remains comparable, especially when $r$ and $w$ are close and small.

\subsection{Experimental Analysis}
\begin{figure}[t]
  \centering
  \includegraphics[width=0.9\textwidth]{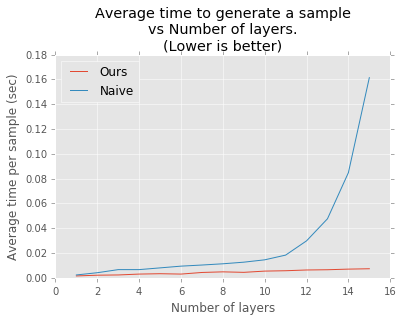}
  \caption{Timing experiments comparing the generation speeds of the na\"{i}ve algorithm and Fast Wavenet.}
  \label{fig:timing_experiments}
\end{figure}

We will now compare the speed of our proposed implementation with the na\"{i}ve implementation. In Figure \ref{fig:timing_experiments}, we generated samples from a model containing 2 blocks of $L$ layers each, using the previous implementation and ours. Results are averaged over 100 repeats. When L is small, the na\"{i}ve implementation performs better than expected due to GPU parallelization of the convolution operations. However, when $L$ is large, our efficient implementation starts to significantly outperform the na\"{i}ve method.

%%%%%%%%%%%%%%%%%%%%%%%%%%%%%%%%%%%%%%%%%%%%%%%%%%%%
\section{Conclusions}
In this work, we presented Fast Wavenet, an implementation of the Wavenet generation process that greatly reduces computational complexity without sacrificing space complexity. The same fast generation scheme can be applied anytime one wants to perform auto-regressive generation or online prediction using a model with dilated convolution layers. The authors hope that readers will find the algorithm useful in their future research.

%%%%%%%%%%%%%%%%%%%%%%%%%%%%%%%%%%%%%%%%%%%%%%%%%%%%
\subsubsection*{Acknowledgments}
Authors would like to thank Wei Han and Yuchen Fan for their insightful discussions.

%%%%%%%%%%%%%%%%%%%%%%%%%%%%%%%%%%%%%%%%%%%%%%%%%%%%
\bibliographystyle{iclr2017_workshop}
%\bibliography{fast_wavenet}
\bibliography{iclr2017_workshop}

\end{document}